\documentclass[man,floatsintext,draftall,natbib]{apa6}

\makeatletter 

\usepackage[english]{babel}
\usepackage[utf8x]{inputenc}
\usepackage{amsmath}
\usepackage{graphicx}
\usepackage[colorinlistoftodos]{todonotes}

\usepackage{soul}
\usepackage{csquotes}
\usepackage[breaklinks]{hyperref}
\hypersetup{
     colorlinks=true,
     linkcolor=violet,
     filecolor=green,
     urlcolor=blue,
     citecolor=red,
}

\newcommand\eatpunct[3]{}
\makeatother 

\title{Everyone's Universe: Teaching Astronomy in Community Colleges}
\shorttitle{Teaching in Community Colleges}
\author{\href{mailto:rfrench@miracosta.edu?subject=IoP eBook AE vol 2}{Rica Sirbaugh French}}
\affiliation{Department of Physical Sciences, \href{http://www.miracosta.edu}{MiraCosta College}, One Barnard Drive, Oceanside, CA 92056}

\abstract{Despite the negative stereotypes still overshadowing community colleges, scores of freshmen nationwide are deliberately beginning their college careers at these institutions and the numbers are increasing more than twice as fast as those of the four-year schools. Approximately 300,000 of these students take introductory astronomy each year as the \textit{last formal exposure to science} most of them will ever have, and at least one-third of these students do so at a community or two-year college. The importance of investing in and devoting resources and training to serve this population -- \textit{everyone}, demographically speaking -- cannot be understated. Yet the overwhelming majority of those who \textit{do} serve this population are lacking in both areas. The community colleges' heavy emphasis on teaching and student success creates both challenges and opportunities that educators must meet head-on using a variety of methods and innovative strategies, teamwork and faculty support systems, and clever workarounds. Here, we introduce both the student and faculty populations, examine the variables affecting both populations, and offer some advice for those looking to teach introductory astronomy at a community college.}

\keywords{Astro 101, astronomy, college, community college, education, education research, faculty, general education, non-science majors, physics, professional development, science, STEM, students, teaching, university}

\begin{document}

\maketitle

\tableofcontents

\clearpage

\section{Introduction}
\label{intro}

Today’s society propagates some deeply entrenched and widely varying ideas about what community colleges are and are not. Most folks will readily share their thoughts on everything from the quality of instruction and how two-year schools differ from four-year colleges and universities, to the contrasting student populations and even the qualifications of faculty and administrators. Let’s face it: it’s hard to argue that there isn’t still a lingering stigma attached to teaching, well, almost \textit{any} subject, at a community college. Even seasoned educators are sometimes not immune to this nor do they always have the right information themselves! It \textit{is} everyone's universe and, unsurprisingly, we all have opinions. Whatever those current opinions are, the data are robust and there is plenty of evidence to illuminate the true picture. While I have taught both physics and astronomy at large four-year institutions, I’ve been at a community college teaching primarily astronomy and the occasional physics or physical science class since 2004. For this writing, I also asked colleagues at other community colleges about what they thought should be included here. Their insights, whether on hiring practices or teaching the universe to everyone, are simultaneously predictable and revealing. So let us go straight to the sources, take stock of the situation, and have a look through the lenses of some accomplished practitioners.

\subsection{Learning Outcomes}
\label{outcomes}

By the end of this chapter, the reader will be able to:
\begin{itemize}
    \item describe the demographics and general characteristics of the population taking Astro 101;
    \item articulate the importance of this population and the implications with respect to science literacy;
    \item compare and contrast the challenges facing community college students with those at the four-year institutions;
    \item assess his/her own level of preparation to teach astronomy, particularly in the community college setting, and explain why it matters who teaches the introductory courses;
    \item differentiate teaching at a community college from doing so at a four-year college or university;
    \item evaluate his/her own commitment to teaching as a profession;
    \item formulate a plan for his/her professional development;
    \item examine and develop strategies and skills for strengthening job applications and interviewing for community college faculty positions; and
    \item decide whether teaching at a community college might a good career choice for him/her.
\end{itemize}

\section{Why It Matters: Get to Know the Players}
\label{whyitmatters}

American community colleges have long been viewed as the ``last resort'' or even the ``land of infinite second chances,'' with the negative connotation often referring to both students \textit{and} faculty. But why? The very nature and design of the junior college system was, originally, to (1) provide lower-division general education [to everyone], regardless of whether s/he might want to transfer to a four-year institution for more specialized study, and (2) offer technical and vocational training [to everyone] to support our nation's changing workforce demands \citep{tycguidelines}. The federal government championed the public two-year institutions, advocating for the diversification and expansion of these educational opportunities for all Americans. Today, the combined system of community colleges in the U.S. has evolved into a massive establishment that is now home to the most accessible and arguably most teaching-focused arrangement of post-secondary educational institutions in the nation. Indeed, an ever-increasing fraction of \textit{all} college-bound students are choosing to take courses at community colleges before transferring to four-year institutions, all while this negative stereotype inexplicably continues to propagate. Their reasons are myriad and not centered simply on lower costs or the misplaced notion that classes at two-year schools aren't as rigorous. Rather, these students are deliberately choosing to enjoy significantly smaller class sizes and more personal attention, a greater sense of community, stronger student support infrastructures, greater availability of general education courses, more flexible schedules, and even geographic convenience.

\subsection{Enrollments}
\label{enrollments}

\subsubsection{The bigger picture}
\label{bigpicture}

Estimates vary but it’s safe to say that between one-third and one-half of all undergraduates in the United States spend at least part of their freshman year enrolled in a community college \citep{smithetal2004, nces2017-144}. The U.S. Department of Education’s \href{https://nces.ed.gov/}{National Center for Educational Statistics} (NCES) projects enrollments in two-year institutions to grow 21 percent between 2015 and 2026, from 6.5 million to 7.8 million while that of the four-year institutions increases by only nine percent to 11.7 million \citep{nces2017-144}. If that doesn’t already have your attention, look at the numbers specific to astronomy.

\subsubsection{Astronomy}
\label{astro}

\cite{tucker1996} speculated that the majority of all college students taking introductory astronomy in the United States probably do so at two-year institutions. In the early 2000s, estimates put the number of students taking introductory astronomy in the U.S. at approximately 250,000 each year, with 40 to 50 percent of them doing so at community and other small colleges without astronomy or physics degree programs of bachelor's or higher \citep{fraknoi2001, fraknoi2004}. Though there is evidence that slightly more students are now taking introductory astronomy, the fraction that does so in a community college likely remains between one-third and one-half.

The \href{http://www.aip.org/}{American Institute of Physics} (AIP) publishes \href{https://www.aip.org/statistics}{statistics} on introductory astronomy enrollments at institutions granting at least bachelor’s degrees. During the 2015-16 academic year there were 750 physics departments \citep{nmphysdepts2017} and 81 astronomy departments \citep{nmastrodepts2017}, 39 of which are completely separate from their physics counterparts while the other 42 are combined physics and astronomy departments. Departments reported “First Term Introductory Course Enrollments” where they were instructed to include only students taking their first term of a stand-alone (i.e. not a continuation of a sequence), entry-level course. Those in the 750 physics departments included both astronomy and physical science courses because they typically have a significant astronomy component. Table~\ref{table:enrollments} shows that year’s introductory astronomy enrollments.

\begin{table}
\centering
\begin{tabular}{rr}
Astronomy and combined departments: & 54,056 \\
Physics departments\tabfnm{*}: & 143,020 \\\hline
Total: & 197,076
\end{tabular}
\caption{\label{table:enrollments}Introductory astronomy enrollments for 2015-16 compiled from AIP statistics on four-year astronomy \citep{nmastrodepts2017} and physics \citep{nmphysdepts2017} degree-granting institutions.}
\tabfnt{*}{While these enrollments include both introductory astronomy and physical science courses, they do not include conceptual physics courses, as introductory physics enrollments are tallied separately in the physics report.}
\end{table}

The 81 astronomy departments surveyed by AIP are located at institutions also represented in the list of 750 physics departments surveyed. This means the nearly 200,000 students in the AIP group represent about 25\% of the 3011 four-year colleges and universities counted in 2014-15 by NCES (the latest data available at the time of this writing; \citealp{nces2017-094}). We do not currently have information on introductory astronomy enrollments in the other 75\% of four-year institutions (those without astronomy or physics departments).

Using data from 2011, AIP estimates that 71\% of the two-year institutions offering physics courses also offered astronomy courses, resulting in approximately 51,000 students taking introductory astronomy at those institutions \citep{wcphysenroll2013}. Since, as AIP points out, this is nearly equivalent to the 52,000 introductory astronomy enrollments in astronomy degree-granting (four-year) institutions in 2010, it is logical to assume a similar parallel for the 2015-16 data. This adds another 54,000 or so students taking introductory astronomy at the 1616 community colleges and other two-year institutions counted in the NCES data \citep{nces2017-094}. Note, however, that this figure does not include enrollments in two-year institutions that offer astronomy but not physics. Therefore Fraknoi’s \citeyearpar{fraknoi2001,fraknoi2004} estimates of 100,000 to 125,000 – which are largely derived from self-reported data from those teaching introductory astronomy at community and other two-year colleges – are likely still reasonable.

Thus there are \textit{at least} 300,000 students taking introductory astronomy annually in the United States. This is an astounding \textit{10\% of incoming freshman every year} \citep{nces2017-094} and, as near as we can tell, 33 to 42 percent of them are taking that introductory astronomy course (hereafter referred to as “Astro 101”) at a community college.

\subsection{Who Takes Astro 101?}
\label{whoarethey}

\subsubsection{Everybody}
\label{everybody}
Seriously. Arguments about how the Astro 101 populations in the two- and four-year institutions are demographically different in significant ways simply do not hold up. Granted, there are some practical considerations that are much more prevalent in the two-year schools’ populations (addressed later) but those do not impact the general demographics describing who takes Astro 101.

\subsubsection{The general education population: non-science majors}
\label{GEpopnonsci}
The U.S. turns out less than 500 astronomy majors each year \citep{nmastrodepts2017} so it’s pretty obvious that students taking Astro 101 aren’t astronomy majors. Years ago it wasn’t a stretch to assume that the overwhelming majority were probably not science majors at all. But now we don’t have to assume; we can show this is definitely the case. Two-thirds or more of students taking Astro 101 report a major or area of interest in a non-STEM field \citep{rudolphetal2010, deminghufnagel2001}. Science majors notwithstanding, these students are the future: they will become journalists, healthcare workers, CEOs, tradespeople, politicians, law enforcement officers, humanitarians, attorneys, parents, and fellow citizens. \textit{That introductory astronomy course could be the only college-level science course many of these folks will ever take} \citep{partridgegreenstein2003}. And – wait for it – \textit{they’re also the next generation of K-12 teachers!}

As many as 40\% of those taking any college-level introductory science course indicate they plan to become licensed teachers \citep{lawrenz2005} and in the study by \cite{rudolphetal2010}, 25\% of introductory astronomy students self-identified as education majors. Given the numbers, this represents an almost unprecedented opportunity to help shape our nation’s scientific literacy and attitudes about science and its roles in society for literally generations to come \citep{rudolphetal2010,pratherrudbriss2009,fraknoi2001}. In the words of  \cite{rudolphetal2010}, “...we can think of our Astro 101 courses as...professional development courses for future teachers.” Indeed, it is perhaps their \textit{first} professional development experience!

If ever there was an argument for elevating a general education science course to top-tier status with the best trained faculty and enormous resources, this might well be it. Our experiences with these people in \textit{the introductory astronomy classroom could be their last formal encounter with science in their entire lives} – no matter where you teach it! No pressure, right? Our potential for making lasting positive impacts on the voting public, both now and for the foreseeable future is, frankly, staggering.

\subsubsection{Student demographics}
\label{studentdemog}
One of the most wonderful things about teaching introductory astronomy anywhere is recognizing that your class is truly a representative cross-section of humans in America \citep{rudolphetal2010,deminghufnagel2001}. They are of every race and ethnicity and span the range of socioeconomic classes and academic abilities. Just over 80\% come to us through the American public K-12 education system and about one-third have a parent whose education went beyond high school. Roughly 90\% have never taken astronomy before and slightly more women than men take Astro 101. \cite{rudolphetal2010} asked 15 different demographic questions of nearly 2000 students enrolled in 69 sections of Astro 101 spread across 31 U.S. institutions. The work of \cite{deminghufnagel2001} derives from 3800 students responding to 12 student background questions in 66 different Astro 101 classes nationwide. Both included all kinds of educational institutions, illustrating that Astro 101 everywhere is a genuine melting pot in terms of almost every demographic you can think of.

\subsubsection[Are they ready for college?]{Are they ready for college?\eatpunct[]}
\label{readyforcollege}
Apparently not. According to \cite{ACT2017}, only 37\% of 2017’s high school graduates meet the college performance benchmark in science and only 41\% do so in math. At the time of this writing, the \cite{nationsreportcard} reports even more dismal statistics: 22 and 25 percent, of 12th graders place “at or above ‘Proficient’” in science and math, respectively. Perhaps somewhat surprisingly, it is students’ \textit{preparation in high school mathematics -- not science (!) -- courses that correlates more strongly with success in college science courses} \citep{sadlertai2007}. A science course taken in high school will almost always positively impact a college-level science course \textit{in that same discipline}, but generally speaking, more high school science does \textit{not} translate to better performance in college-level science across the board. It’s the mathematics preparation that matters most. In fact, the more math courses and more advanced the math courses taken, the higher the rates of overall college persistence and completion \citep{phillippetekle2016}.

\cite{ACT2017} also reports that only 47\% of 2017’s high school diplomates read at a college level while not even two-thirds (61\%) meet the college performance benchmark in English. According to the \cite{nationsreportcard} only 37\% of 12th graders are “at or above ‘Proficient’” in reading and only 27\% in writing. \cite{completecollege} indicates that of the students enrolling in two-year colleges for the first time, 34\% must take remedial courses in English and 52\% must do so in math. Combined with the 12 and 24 percent, respectively, that require remediation in the four-year schools, that is an incredible number of \textit{students who are not ready for and cannot take college-level English or mathematics courses} despite being able to enroll in college. \textit{Yet they \underline{can} enroll in your Astro 101 course!} Because they are introductory general education courses, virtually none of the introductory astronomy courses at any institution have prerequisites. Remember, too, that most two-year institutions are “open enrollment” institutions meaning there are no academic entrance requirements. In fact, \textit{some students enrolled at community colleges have neither high school diplomas nor GEDs but can still enroll in Astro 101}. While this may sound like a horrific rabbit hole to some, there is actually a strangely bright light at the end of this particular tunnel.

Perhaps one of the most surprising – and encouraging – revelations came when \cite{pratherrudbrissschling2009} showed that \textit{all of these students, regardless of the institution type, class size, or their backgrounds, are capable of achieving similar learning gains in Astro 101} (note this is part of the same study as the \citealt{rudolphetal2010} data). For what feels like eons, instructors have argued to the proverbial death -- using little to no evidence -- that this wasn’t possible, that the four-year institutions have a distinct advantage since their students are “filtered” through academic entrance requirements. Admittedly, it sounded like a plausible contention. The idea that one could actually conduct legitimate education research using his or her own students and that this could provide the data to either refute or support those claims initially flew completely under the radar. But savvy instructors at all kinds of institutions watched, learned, experimented, recorded, tweaked, and reported bits and pieces of what did and didn’t work in their classrooms over the years as learner-centered teaching strategies (see Chapter 1 of this volume) became more and more prevalent. Discipline-based education research (DBER) gradually gained traction before quickly becoming all the rage. The notion that education research could -- and \textit{should} -- inform what we do and how we do it in our classrooms, just as our discipline content research informs our science, finally began to take root. Physics education research (PER) led the way for most of us with astronomy education research (AER) soon to follow, finally garnering some game-changing attention somewhere in the last ten to fifteen years. (See other chapters in this volume, particularly 1, 3, 4, and 9.)

And now we have the evidence. They may not all be ready for college-level work, but interactive learning methods are capable of helping all students, regardless of academic preparedness or demographics. In fact, \cite{freemanetal2014} meta-analyzed over 200 studies of various active learning strategies used in a variety of undergraduate STEM courses and concluded that the techniques appear generally effective across all STEM disciplines and for all class sizes. Many instructors in the \cite{fraknoi2004} data even indicated that they developed a strong sense of pride when “...students of diverse backgrounds and abilities were able to succeed in their astronomy class.” So never fear: a thoughtfully designed and well-implemented learner-centered Astro 101 course (see relevant chapters in this volume, e.g. 1-4) has the potential to bring everyone closer together, both in terms of social constructs and academic performance \citep{rudolphetal2010,pratherrudbrissschling2009}.

\subsection{Special Considerations}
\label{specialconsid}
Along with these interesting and refreshingly diverse perspectives comes a set of special considerations that, while certainly not absent from the four-year institutions’ populations, are a way of life for the students taking courses at two-year institutions. While all institutions suffer with a contingent of enrollees consistently unprepared for class meetings, those with teaching experience in community colleges know that for their students, it is these practical considerations, rather than laziness, that are most often the culprits. How much do you hold their hands? How much do you hold them responsible for? Is it really our responsibility to “meet them in the middle” as so many administrators keep preaching? An instructor’s empathy and sense of compassion wage never-ending battles with the need for structure, rules, and deadlines. So the culture of the two-year school morphs to adapt. You can read about it, hear about it, and even witness it yourself when you visit another instructor’s class. But until you’ve been in front of those students yourself and interacted with them class after class as their instructor, you don’t really know that audience. Until you’ve experienced it firsthand and dealt with their specific kinds of issues, you don’t truly understand it \citetext{C. Hirano, personal communication, March 20, 2018}.

\subsubsection{Time [not] spent on campus}
\label{timenotspent}
While three-quarters of students at four-year institutions are enrolled full-time, only about one-third of community college students attend full-time \citep{nces2017-094,phillippetekle2018}. An estimated 28\% of two-year institutions are residential campuses \citep{phillippetekle2018} but most of us teaching Astro 101 at “traditional” community colleges are not doing so at a residential institution. Numerous community college faculty (myself included) anecdotally cite the combination of an overwhelmingly part-time student population with a non-residential campus as the primary reason that few students avail themselves of the numerous support systems in place, from individual instructors’ office hours and help sessions to the typically robust tutoring, testing, counseling, and other support services offered. These students simply do not spend any more time on campus than they have to and it’s almost certainly a function of the other demands for their time. It isn’t that they \textit{refuse} to devote more time to their studies; they usually feel they have no choice and are forced to sacrifice school and study time to meet life's other demands.

\subsubsection{Life, family, and culture}
\label{life}
A big part of every college student’s life is trying to figure out a reasonable school-work-life balance. But the average age of community college students is 28 \citep{phillippetekle2018} so it’s a safe bet that more of these students are already dealing with the responsibilities of raising families of their own. Most of them are holding down either full-time jobs or multiple part-time jobs and nearly one-fifth of community college students are single parents \citep{phillippetekle2018}. The various forms of “life happens,” e.g. childcare snafus, car troubles, sick kids, getting called in to work unexpectedly, etc., are a never-ending struggle among this population. The older “non-traditional” students, who may be working on second careers or coming back for the degree they never finished as a twenty-something, are not immune either as they are often juggling life’s hang-ups while caring for aging parents and helping support their own college-age children.

Don’t forget, too, that compared to most four-year institutions, community colleges often have larger fractions of students who are first generation college students, veterans reintegrating into civilian life, and previously incarcerated individuals striving for a fresh start. There are also more instances of two-year schools located in regions with much narrower and very specific demographics such as tribal colleges or campuses in areas experiencing critical economical distress where most residents are below the local poverty level.

All of these students bring different cultural norms and sets of circumstances with them to school each day. Figuring out how to embrace it all and turn the challenges into strengths and opportunities is one of the hallmarks of teaching both Astro 101 and teaching at a community college.

\subsubsection[No filter, no problem. \textit{Wait! Problem...}]{No filter, no problem. \textit{Wait! Problem...}\eatpunct[]}
\label{filter}
Students at four-year institutions are “filtered,” having jumped through a tedious series of admissions hoops that includes minimum academic achievement standards. In short, they know how to be students. That said, their study habits aren’t necessarily any better than those of the “unfiltered masses” at the community college. Rereading, highlighting or underlining, and “taking notes” [read: “copying”] are still the default study tactics despite years of research indicating these methods are largely ineffective \citep{roediger2013}. We must guide all of our students, regardless of academic ability or preparedness level, through the kinds of activities that are shown to actually facilitate real learning (e.g Chapters 1-3 in this volume).

The fact that many community college students generally do not know how to be students, turns out, is a pretty big deal. The Filtered Ones have amassed and refined a set of skills that many students in the two-year schools have not. They know what it means to set a schedule, organize tasks, acknowledge deadlines, manage their time, and seek out help with respect to their studies – and this is very different from employing those same tasks in other aspects of life. Whether they are good at it or not is an entirely different proposition. The point is they are at least \textit{aware} that these skills are essential to their academic success. Most students in community colleges either don’t yet have this skill set or have not practiced most of it in recent history so it is long forgotten. Again, they can sometimes manage other aspects of their lives well enough with the analogous skills. But add academic studies into the mix and things can go sideways pretty quickly.

We end up helping them learn the intricacies of post-secondary education, navigate the bureaucracies that we and our administrations put in place, and re-teaching those same life skills as they apply to education. So much so that these curiously-named “high-impact practices” now form the basis for many courses, orientations, and entire engagement programs developed and implemented in many community colleges across the nation \citep{cccsedegrees2013}. Now we just have to make these programs and courses integrated, required components of the community college experience. To a college student, “optional” translates to “not gonna happen,” so in order to make a difference, engaging community college students must be a deliberate and focused effort. While much of this is out of the hands of most astronomy instructors (such as first year experience and orientation programs), others such as tutoring, learning communities, supplemental instruction, experiential learning, and early intervention (ref. Chapters 1 and 2 in this volume) can be effectively integrated into our introductory astronomy curriculum.

\subsection{Who Teaches Astro 101?}
\label{whosteaching}

\subsubsection[Not astronomers!]{Not astronomers!\eatpunct[]}
\label{notastronomers}
No kidding. According to \cite{fraknoi2004} only 23\% of introductory astronomy instructors in community colleges have degrees in astronomy and including the four-year schools barely raises that figure to 25\%. \cite{fraknoi2004} and \cite{tucker1996} show that most introductory astronomy instructors have their degrees in physics (you might be one!). But before you hand-wave that away as perfectly acceptable, consider it from this perspective. The enrollment data at the beginning of this chapter show that 48\% of students taking Astro 101 do so in a pure physics department and another 40 to 50 percent take the course in a department that does not offer a degree in \textit{either} physics or astronomy. This means that \textit{88\% to 98\% of these students are taking introductory astronomy with instructors who, themselves, very likely have little to no formal training in astronomy}. If your own degree is in physics ask yourself this, “How much astronomy-specific content did I get while fulfilling my physics degree requirements?” If you’ve ever been one of these instructors – you’ve taught introductory astronomy with only a cursory prior exposure to it – how would you answer this next question (be honest!): “Knowing what you now do, would you say you were truly qualified to teach \textit{astronomy}?” Answers ranging from a rather unconvincing “sure” and an uneasy “not really” to a resounding “definitely not!” are common from instructors in the professional development workshops I’ve co-facilitated\footnote{One of the more snarky (and terrifying) analogies sounds something like this. ``Oh, by the way...your helicopter pilot has no real training in rotor-wing aircraft. No, no, it's okay. Really. She's more than qualified to fly \textit{fixed-wing} aircraft. You know, \textit{airplanes}? And that's enough. There's no need for special training just because it's a different type of aircraft.''}. Having a degree in physics myself I can say that my answer to the first question would have been “virtually none,'' had I not enrolled in an ``unnecessary'' Astro 101 course simply to maintain full-time enrollment status one term. A subsequent degree in astronomy made all the difference for me personally but had it instead been in physics, I’m sure my answer to the second question would have been “definitely not!”

To be clear, no one is questioning whether a physicist is \textit{capable} of teaching astronomy. There is little doubt he or she most certainly is, obviously having more than adequate academic preparation in the necessary physics. The question is whether those without any experience in the \textit{astronomy content} should be doing so given the enormous implications of the introductory astronomy survey course (see ``\nameref{GEpopnonsci}''). The state of California thinks so. So does the \href{http://aapt.org/}{American Association of Physics Teachers} (AAPT) --  even while many of those very instructors see itl differently.

Those who have taught introductory astronomy know that it is much more than the basic phenomena of Newton’s laws, gravity, light, and optics we all learned in “pure” physics courses. But the state of California considers physics and astronomy to be essentially the same thing, a single “discipline” or “area.” The discipline is labelled “Physics/Astronomy” and in order to teach either subject at a community college in California, one must meet the following minimum qualifications: “Master’s in physics, astronomy, or astrophysics OR Bachelor’s in physics or astronomy AND Master’s in engineering, mathematics, meteorology, or geophysics OR the equivalent” \citep{woodyardCCC2017}. The “equivalent” loophole gives individual colleges and districts some leeway in determining what is appropriate for their courses since it allows for things like professional experience commensurate with an earned degree or degree titles that don’t \underline{exactly} match (not kidding here) any of those in the list. For example, unless the school or district has local governance policies in place that allow variations in degree titles, an applicant with a bachelor’s in astronomy and master's in “engineering physics'' does not meet the minimum qualifications and therefore must go through the equivalency process. Typically, an interdisciplinary faculty committee considers equivalency applicants very carefully, examining their academic preparation and professional experience closely before making a recommendation that is passed up the administrative chain, subject to further scrutiny and recommendations. Some other states have similar such minimum qualifications or variations on them, making it easy to recognize a qualifying combination of degrees that could exclude astronomy-specific content entirely. A few such examples are specifically called out in the \cite{fraknoi2004} survey comments, highlighting both this issue and that of a general sense of isolation (discussed in later sections).

\begin{displayquote}
“One instructor with a master’s degree in math wrote, ‘I actually taught two semesters of astronomy without ever taking an astronomy course myself!’ (Whether this is a good or bad thing for the world is left as an exercise to the reader).”

“I wish I had someone to ask questions of. I am a physicist, not an astronomer.”

“I am always in need of advice...since I have a PhD in psychology and not the physical sciences.”
\end{displayquote}

In the early 2000s this quandary surfaced enough times both in inquiries from administrators and in the accreditation proceedings of two-year colleges across the nation that the Executive Officer of the AAPT at the time issued a statement \citep{khoury2004} citing two main lines of evidence to support the position. Specifically (1) “The introductory astronomy course is a ‘science literacy’ course, designed to present a broad background in the subject typically to non-science-oriented students and not for future astronomy ‘majors,’” and (2) the basics are “all topics included in a physics teacher’s background” because of the “strong overlap...in the curriculum and skills required in the two areas” and as evidenced by the number of universities with unified departments of physics and astronomy. Some time later, the AAPT Executive Board endorsed and posted a subsequent statement \citep{AAPTBoard} supporting an “emphatic ‘\textbf{Yes}’” in response to the question, “Does a degree in physics qualify a person to teach introductory astronomy at the collegiate level?” The Board’s position rests on three tenets: “...the nature of the curriculum in the two fields of physics and astronomy...on common practice regarding how introductory astronomy is offered across the United States; and...on the role of the introductory astronomy class in the college curriculum.” Though this statement largely reconstitutes the article written by the former Executive Officer, the Board does go a bit beyond these superficial arguments by pointedly calling out practical experience (such as working in a planetarium, participating in research, and attending workshops offered by professional societies) as a “sufficient” qualifier. It also includes the AAPT Space Science Committee’s recommendation to evaluate potential candidates by examining both one’s broad course preparation and relevant work experience.

While all this sounds well and good, the evidence clearly shows that the argument for Astro 101 being an introductory course that serves ``non-science-oriented students'' (see ``\nameref{GEpopnonsci}'') supports the position in diametric opposition to the one claimed. Indeed, a case could be made that the data argue even more strongly for astronomy instructors to be trained not only in the astronomy-specific content, but also in pedagogy and particularly communicating science to non-scientists! Additionally, the argument about significant overlap in curriculum appears to be, at best, a highly variable function of an institution's program design, and at worst, simply false.

Curiously, the \href{http://aas.org/}{American Astronomical Society} (AAS) seemingly has no official position on the matter. However, in a 2016 report authored by the \href{https://aas.org/education/aas-education-task-force}{AAS Education Task Force} the dilemma is specifically called out: “The AAS needs to be mindful of the fact that most Astro 101 instructors are not research scientists in astronomy and may lack the ability to teach – or at least be uncomfortable with teaching – some Astro 101 topics'' \citep{AASreport2016}.

\subsubsection[It matters!]{It matters!\eatpunct[]}
\label{itmatters}
It \textit{does} matter who teaches introductory astronomy. The \cite{cccseparttime2014} reports that roughly 58\% of all courses at community colleges are taught by part-time faculty. Though there are notable exceptions, most part-time faculty have few ties to a given institution and can be less invested in the undertaking. (There is some evidence to suggest this may be more the fault of the institutions themselves rather than the instructors; more on this later.) \textit{The connection between Astro 101 and whatever comes next for these students is important -- and hinges critically on the experience with that instructor}, a point that the \href{https://www.chronicle.com/}{Chronicle of Higher Education} very recently saw fit to call out \citep{supiano2018}. Regardless of whether the instructor is full-time or part time, if the students have a bad experience, it can initiate a butterfly effect. Most notably, (1) community college students taking remedial courses from part-time faculty are less likely to persist to the next course, (2) students taking introductory STEM courses from non-tenure-track faculty in the four-year schools are 1.5\% more likely to switch to a non-STEM field, and (3) the greater the percentage of non-tenure-track faculty teaching at the four-year schools, the less likely  students are to graduate \citep{supiano2018}. Remember how many students take Astro 101? Remember how that course could be their last experience ever with formal science? Remember who they are and what they represent? (See ``\nameref{GEpopnonsci}.'') And while it is true that we have few STEM majors in our Astro 101 classes, the propensity of such a course to draw bright, undecided students into such fields is not at all negligible.

\subsubsection{Preparation and training}
\label{prep+train}
The Physics vs. Astronomy Heavyweight Championship aside, most instructors are generally well-educated with 94\% of community college instructors having advanced degrees \citep{fraknoi2004}: 31\% have doctorates and 63\% have master’s degrees. The 93\% of four-year college instructors with advance degrees includes 76\% doctorates and 17\% master’s degrees. All this education, yet a large fraction of STEM graduate students receive no formal training whatsoever in pedagogy or virtually any aspect of teaching or mentoring before stepping into a classroom – which most of us will do at least once during our careers whether we want to or not. Nearly all AAS members who responded to the \citet{AASreport2016} Education Task Force’s survey (93\%) indicated that they teach (or have taught) at least one class of some kind and almost three-quarters of respondents teach Astro 101. Approximately one-third of respondents said they received no training whatsoever with regard to teaching and almost 70\% indicated no training in mentorship of any kind.

So what is the aspiring astronomy instructor to do? Professional societies such as the AAS and \href{http://astrosociety.org/}{Astronomical Society of the Pacific} (ASP) regularly offer opportunities to learn about both mentoring and teaching practices, often with reduced registration fees for community college and other local educators. The AAPT is exclusively focused on teaching and the \href{http://astronomy101.jpl.nasa.gov/}{Center for Astronomy Education} (CAE) is dedicated specifically to improving the teaching and learning of introductory astronomy, Earth, and space science. (More on professional development later.)

\section{You Matter: The Job of Community College Faculty [in Astronomy]}
\label{teachastrocc}

\subsection{Environment, Workload, and Resources}
\label{environment}
In the two-year schools (and even many smaller liberal arts and teaching-focused four-year institutions), astronomy is one of many disciplines in a multi-discipline department. The most common unit is a variation of a physical sciences department that includes disciplines such as physics, geology, earth science, oceanography, and physical science. It is not unusual for chemistry to be included and even things like geography and meteorology are surprisingly common. A full-time astronomy instructor is almost certainly the only astronomer in the department and, more often than you'd think, also the only physicist. In fact, it’s usually the other way around: teaching physics comes first with the astronomy following later, sometimes by choice but sometimes because s/he was asked to take it on. In some cases, one is hired specifically as a full-time astronomy instructor but either way, it is typical for the faculty member to teach more than just astronomy. Physics and physical science are probably most common.

The AIP has only recently added astronomy teaching to their survey of physics faculty in two-year colleges so longitudinal comparisons aren’t yet possible. It’s also not possible to disentangle the astronomy data from the physics data but it is still painfully clear that astronomers and physicists at these institutions are relatively isolated. Sixty percent of surveyed departments had either no or only a single full-time faculty member teaching astronomy and/or physics \citep{wcphysfac2013} and those lone full-timers are also the folks most likely to be teaching courses in additional disciplines. It’s also now clear that a substantial number of those teaching astronomy at two-year colleges are part-time faculty. The part-timers, however, are much more likely to teach only astronomy and physics classes with only about 20\% of them teaching other subjects \citep{wcphysfac2013}.

Full-time instructors can face several challenges that stem from being the lone discipline expert in a multi-discipline department. Duties that are normally within the purview of a department chair often fall to that lone faculty member if the chair lacks the necessary discipline-specific expertise. It is of course reasonable to expect discipline faculty to, as part of their regular responsibilities, handle occasional issues and advocate for their particular programs and students as necessary. But if all of the administrative business for that discipline eventually ends up on the plate of the lone discipline expert who is not the department chair, this is no longer a non-negligible component of one’s workload. Occasionally the faculty member may receive some compensation in recognition of these additional duties, perhaps along the lines of a “lead instructor” or other similar designation, but most often they do not. When there are not a large number of astronomy classes or students, this may not warrant too much concern. If, on the other hand, the program is larger, has more than a few classes, many students, and several part-time faculty to supervise, the “workload creep” can be substantial.

In contrast to most four-year institutions, many community colleges have little or no instructional support, e.g. lab prep assistants or equipment managers. Though there are exceptions, a lone full-time instructor in the discipline is often responsible for everything, from purchasing equipment and supplies to maintaining that equipment and any associated facilities (e.g. computer lab, observing site, etc.). While it is no longer uncommon to have access to a computer lab, a community college with an actual observatory or planetarium is still an exception. Usually, there are binoculars and/or portable telescopes that must be lugged back and forth between their storage area and a makeshift observing site. More advanced equipment isn’t all that common and is normally seen only at larger institutions with the even more rare advanced classes and corresponding budgets.

Unless you teach at one of those large institutions with the rare complementary budget, money will be tight. Data from \citet{fraknoi2004} show the average annual budget for astronomy programs (not including physics) in community colleges to be around \$940 and including all the surveyed institutions only raises this figure to \$1127. My own program’s budget is right in line with these figures and colleagues elsewhere report similarly. Keep in mind that this amount is usually meant to encompass \textit{everything} needed to keep the program running, from replacing equipment to photocopying costs for all classes.

There is little to no money for conference travel and professional development is frequently limited to in-house workshops and seminars on more universal content like diversity and equity issues, institutional procedures, statewide initiatives, and high-impact practices. Professional development for those teaching astronomy is very specialized so local opportunities are generally rare. Though professional societies sometimes offer grants or reduced fees to community college instructors, a significant travel distance is almost always involved so it is frequently cost-prohibitive. Note, too, that if you are the only astronomer, finding qualified employees that meet the legal requirements to substitute in the classes you'll have to miss is often a sticky subject with department chairs, deans, and other administrators. On a more positive note, more and more external resources have been devoted over the past decade or so to building a robust community of astronomy education professionals, resources, and opportunities so it is gradually becoming easier to transcend isolation and budget limitations \citep{wallerslater2011}.

There are virtually no resources or support for personal research and, frankly, no time to do it anyway. This is okay since virtually no two-year institutions have an expectation that you maintain a research program -- or ever publish again, for that matter. In practice, the labors of love do still occur. Motivated instructors will maintain ties with collaborators elsewhere and sometimes find the time to squeeze in projects that matter to them.

A full-time faculty member at a two-year institution teaches, on average, the equivalent of five three-credit courses per semester, \textit{every} semester. Of course exact loads vary across institutions but so do the courses and their credit values. For many though, a full-time teaching load is the equivalent of 15 credits where each one credit is a “package” of one 50-minute classroom hour and one hour of prep work and grading per week. Thus a full load means 30 hours of instructional time each week: 15 contact hours in class and 15 hours outside of class. Using a 40-hour work week model (I know, don’t laugh – more on that later), the remaining ten hours is comprised of additional student contact time and institutional service duties. At my institution the expected breakdown is five “student hours” and five hours of “collegial governance” duties. Those five weekly student contact hours must include at least two hours of regularly scheduled drop-in office hours while the rest of the time is spent on things like responding to students’ emails and phone calls, taking additional student meetings, and advising. The remaining five “governance” hours are consumed by various committee meetings (departmental, institutional, and/or district), the work for those committees, and any other duties in service of the institution.

Community colleges in particular are big on the concept of “collegial governance” though that has become somewhat of a buzz-phrase in recent years, coming to mean many different things to many different people at many different institutions (go ahead – I dare you to Google it...). Generally speaking, this refers to the practice in which all constituencies participate in a policy of shared governance, each contributing regularly and meaningfully to the functioning and oversight of both the big things (like institution-wide policies and procedures) and the smaller day-to-day operations (like how your department handles prerequisite challenges). As one might expect, the implementations of such a model vary extensively across the nation but as a rule, all full-time faculty at pretty much any institution are expected to participate at some level and it is even written into the job description. Part-time faculty typically have no such obligation to participate but more and more schools and districts are beginning to offer such opportunities, some with accompanying pay, in order to help foster a sense of community and encourage part-timers to become more involved and invested in their institution.

Most of us are all too aware that the 40-hour work week is basically a unicorn – fantastically elusive and ultimately, a mythical creature. For community college instructors the killer is almost always the time spent grading. If you end up doing all or even most of the discipline’s administrative and support duties, it’s even worse. There are, of course, no graduate students so no teaching assistants. And the nature of the two-year school means that any suitable undergraduates you might try to recruit and train aren’t likely to stick around long enough for it to be worth the effort. Even if they did, how will you pay them? (You read about your budget a few paragraphs ago, right?) So grading? Yeah, that’s probably all you. No matter how many students you have. The trade-off is that you can get to know your students much better and tailor your instruction to address issues you'd otherwise miss if you had a grader. That said, it is almost always the bulk of your workload. But there actually may be a couple of possibilities for enlisting the help of qualified undergraduates. Just don’t count on mining the astronomy club – there probably isn’t one\footnote{Only 20\% of all respondents in the \cite{fraknoi2004} study reported having an amateur astronomy group of some kind at their institution. What fraction are in just the two-year schools is unknown but it must be exceedingly small since the number of available students and transient nature of the population causes high turnover rates, making it difficult for such a group to persist.}. Nevertheless, if you’re willing to put forth the time and effort, there may be some reasonable prospects.

If a student receives federal or state financial aid, he or she might be eligible for work-study funds. If the school does any type of astronomy outreach (like star parties) there is often an argument to be made for hiring a student worker. If there aren’t enough outreach hours to justify the position, see if combining it with a few hours of teaching assistant duties might work. It doesn’t have to be all about grading either; you might be able to get a lab or observing assistant out of it. Education majors might be interested because they want to be teachers. Psychology and other social science majors might just be interested in observing other humans’ behaviors and interactions. STEM majors might prefer grading astronomy homework or being your lab assistant to answering phones or being a receptionist in a front office elsewhere on campus. Others might simply be so proud of their own success in your course that your acknowledgment of it and offer to work as a teaching assistant becomes a life-changing point of encouragement for them. You probably need to go through your school’s career center or student employment office for more information on those possibilities. Reach out to the staff in that office. Even if their office can’t help, they might be able to brainstorm other potential solutions. If none of that presents a feasible option, look for the more creative workarounds.

Practically all two-year schools have some type of formal course that students can enroll in to earn college credit for internships and/or discipline-related work experience. This can take many forms and is certainly not limited to students wanting to “do astronomy” as a career (see the examples in the previous paragraph). The only real hiccups with this plan are likely to be those caused by the fact that the student has to enroll in and pay for an additional credit course. Depending on the individual student’s situation, there may be problems with credit limitations or number of work hours allowed and even financial aid ramifications. Talk with the student and if necessary, get appropriate career, academic, and/or financial aid counselors involved in that conversation.

Another possibility might be service learning credit. If the school has a program for students to earn service learning credits there may be a way for this to qualify. Not everyone understands just how valuable experience like this can be on a resume or CV and how it often goes far beyond the astronomy content or the seemingly superficial notion of assigning grades. Gather the facts, think it through, and promote it appropriately.

Whatever the case, don’t be shy about advocating for the student and the work itself. Leave no stone unturned. Talk with your department chair and dean, too. Even if it doesn’t pan out, you’re probably no worse off than when you started, \textit{the student(s) will never forget that you tried}, and you might have gleaned information that helps you formulate a better plan for making the case next time. Each individual student’s situation is different but who knows? You might end up learning what you need to develop a pilot program that could evolve into a more formal and sustainable solution.

\subsection{Compensation and Benefits}
\label{comp+benefits}
As one might expect, two-year college instructors are generally paid less than their counterparts at the four-year institutions. But like any field, there are exceptions to the rule and they most often occur in geographic regions with high property taxes and costs of living. Many two-year institutions make use of the “industry standard” academic ranking structure for full-time instructors but a few – like my own – do not. In these cases, the pay structure is frequently based on only two factors: one’s level of educational preparation (highest degree and any additional credits) and number of years of classroom teaching experience. Of course variations exist, particularly if the institution implements an academic rank and reward structure. There are procedures in place for advancing on the salary schedule should you later earn additional credits and/or another degree. Sometimes there are fringe benefits to be had such as tuition reimbursement. But...do your homework! You wouldn’t want to be denied advancement on the salary schedule because that institution has a policy stating that credits paid for by that institution cannot then be used for advancement on their own salary schedule.

It is also not uncommon for full-time faculty to take on overload teaching assignments. There are various reasons for this, not the least of which is simply to make up for a deficit in income. The pay for teaching overload classes as a salaried employee also varies considerably across institutions. Some calculate an hourly rate based on one’s full-time compensation, meaning you get paid nearly the same for an overload class assignment as you do for a regular class assignment. Other institutions default to the part-time salary schedule (see below) for any overload assignments, usually meaning that pay is substantially less. And, as one might imagine, overload assignments can impact one’s retirement and leave balances. They even occasionally come with additional responsibilities if that college attaches institutional service or professional development obligations to the number of teaching hours one takes on in a term.

The salary schedule for part-time faculty usually has a similar structure to account for the same variables (educational preparation and teaching experience). The biggest difference is in the dollar amounts. Part-timers are often paid substantially less than their full-time counterparts (\textit{surprise!}). While full-time positions are almost always fully benefitted positions, the benefits afforded to part-time faculty vary wildly from school to school. Some do offer full benefits packages including insurances, sick leave, paid office hours, and the like, but those are the exception and not the rule. Others offer nothing: you’re paid an hourly wage for only the time you spend in class or perhaps a stipend per class per term based on the number of credits and that’s it. Anything else is up to you.

\subsection{Other Part-Time Challenges}
\label{parttimechallenges}
Like many other job markets, full-time faculty positions are greatly outnumbered by part-time ones. Part-time instructors are frequently called ``adjunct professors'' (though my own institutional culture maintains that term has a negative connotation in that it implies inferior worth or otherwise devalues the individual in comparison to a full-time instructor). Many part-time faculty have “regular” jobs elsewhere (industry and research positions in particular are prevalent among part-time faculty in STEM fields) and choose to teach one or more courses, often in the evenings after a full work day, because they love teaching. Countless others, however, stitch together the equivalent of a full-time position by teaching several classes spread across multiple institutions each term. In California these folks are known as the “freeway fliers” because they spend so much time on the road traveling between schools (“freeway faculty” and “roads scholars” are popular monikers in other regions). It is not uncommon for freeway fliers to teach six or seven classes per term (a colleague of mine once taught \textit{nine} classes in a single semester and still occasionally teaches as many as eight!), all while trying to keep straight the policies, procedures, and deadlines for their different institutions.

While full-time faculty at two-year institutions do have dedicated office space they often must share it with others and in some cases the “office” is actually a cubicle. Part-timers, unfortunately, don’t usually fare even that well. Sometimes the rare unattended office is available for first-come, first-served transient use; sometimes there is a designated workspace with supplies; sometimes there’s nothing. If it were only about holding office hours for students there actually wouldn’t be that big of a problem. It is now commonplace for faculty to do so in more inviting spaces, those more conducive to student interaction and collaborative learning. Designated spaces in student services areas, libraries, STEM centers, cafeterias and other places where students typically hang out – even outdoor spaces – are becoming very popular now for holding office hours and help sessions. But that still leaves most part-timers – the majority of teaching faculty – without any private or even semi-private workspace. “My office is the trunk of my car” is a common [not-so-much-a-]joke among part-time faculty everywhere.

\subsection{Rays of Hope}
\label{hope}
If all of this has left you a bit depressed let me assure you that despite the challenges, many of us genuinely love teaching in community colleges and wouldn’t have it any other way. There are \textit{so} many reasons! Here’s just a few.

\subsubsection{Autonomy + academic freedom}
\label{autonomy}
Sure, isolation can sometimes be a problem if you are a single-faculty discipline, but the upside is the number of degrees of freedom you have in virtually all other aspects of the position. The amount of red tape you can cut through is nothing short of a miracle in some cases. For example, you often control the money. How, when, and on what your spend your budget are decisions that \textit{you} make, subject to perhaps a supervisory signature that, depending upon your supervisor's or dean's management style, can sometimes be nearly hassle-free. What textbook should you use? Or should you even use one? Again, all you if you’re a single-faculty discipline. And though there isn’t always significant time to devote to research, the lack of “publish or perish” pressure in a two-year college environment means that you’re “free” to work on whatever you want whenever you want!

Did you know you can teach almost whatever you want in an introductory general education course? Yes, you do have to pay attention to certain guidelines and make sure your course outline of record aligns with articulation and transfer agreements. Yet most don’t even realize that there is a tremendous amount of academic freedom possible, even within institutional or state guidelines and introductory astronomy might just be the most flexible course in the universe (pun intended)!

The beauty of the introductory survey course is that it is rarely, if ever, a prerequisite for anything else, except perhaps the first of a two-course intro sequence (and even then you have two courses with $n-1$ degrees of academic freedom!). Students majoring in a particular discipline don’t customarily take the introductory survey, 101-esque version of the course. \textit{Astro 101 is that course:} generally a single-term “one-off” that majors or minors take only infrequently. So this version is broad enough that there’s a lot of leeway for the instructor to pick and choose what fills in the gaps between the expected larger concepts. Near the beginning of his article on “Teaching at a Community College: Some Personal Observations,” \citet{ball2010} provides an excellent description of this realization and how it manifested in his own adjustment to it teaching introductory history courses. \cite{partridgegreenstein2003} beautifully describe both the context and what it means specifically for an Astro 101 course. In fact, section 2.3 of that seminal work became the basis for what many astronomy educators now refer to as \href{http://tiny.cc/Astro101goals}{“the ‘goals’ document”} distributed by the AAS for many years.

\subsubsection{Teaching as a profession}
\label{profession}
You can now -- and without reservation -- explicitly acknowledge your commitment to becoming a teaching professional. Communities of excellence in teaching practices exist; seek them out – not only in astronomy and STEM fields in general, but also in other disciplines. There may be one or more groups on campus or even a center for teaching excellence or some other such entity. Commit to and immerse yourself in the culture of teaching as a profession. Know that your colleagues and administrators at two-year schools \textit{expect} this of you, often unlike the culture at a lot of four-year institutions. You were immersed in the research culture when you were in graduate school. Become immersed in the \textit{education research} culture now as a teaching professional.

With the emphasis on DBER in recent years, many colleges are renewing their commitments to teaching excellence and throwing resources behind them. It isn’t just lip service anymore to have a department chair, dean, and/or VP that claims to support educational innovations and new technologies. This means you could have even more freedom in the classroom to try new things and still have support even if it doesn’t go well the first time. Many institutions are receiving various state and federal funds to support things like additional “basic skills” initiatives, developing more multidisciplinary learning communities, and even supporting STEM resource centers. Periodic conversations with administrators, initially supportive or not, can have them putting their money where their mouths are more often than you’d think.

This could also mean more money for professional development opportunities. It’s probably obvious by now that lack of time is a significant hindrance to meaningful professional development. But it is also the case that many community colleges require a certain number of professional development hours of their faculty. These obligations are frequently based on the size of one’s teaching load and can be met in a variety of ways. Activities ranging from the usual workshops, seminars, and conferences to simply reading discipline journals and education blogs, and even developing new materials for your courses could satisfy the requirements. Guidelines for what does and does not count as “legitimate” professional development can vary widely among the community colleges so make sure you connect with your college's professional development coordinator or equivalent before you devote a lot of time that you \st{probably} definitely don't have to activities that may or may not count. Then adopt the ``multiple birds with one stone'' mindset: work smarter, not harder.

\subsubsection{Diversity + smaller classes}
\label{diversity}
The magnificent diversity of our students combined with the smaller class sizes \citep{fraknoi2004} frequently lends itself to some unbelievably amazing and bizarre (in a good way) class discussions that go virtually unmatched in most other introductory science classes. Remember, you are one of the Filtered Ones and it’s actually worse than that – \textit{you are in a STEM field} – so interacting with this Astro 101 population is good for you. As my colleague Philip Blanco emphasized, “...it’s different [teaching introductory astronomy] from what you’re probably used to...you probably didn’t get [this] in grad school...so many different kinds of people. It’s so much fun seeing how the different mindsets and personalities interact with the information!” \citetext{personal communication, March 21, 2018}. It isn’t all that unusual for faculty to ruminate on the irony of having a frustrating day at work only to realize that we actually look forward to the Astro 101 classes precisely because of this invigorating potential.

\subsection{Some Advice}
\label{advice}

\subsubsection{Applying + interviewing}
\label{applying}
Before applying for a full-time position at a community college, take the time to do your homework. The two-year schools certainly aren’t for everyone so how could you know? Take a read through these articles by Rob Jenkins, a frequent and popular author on the topic of teaching in community colleges. One gives a good summary of why you should bother to apply \citep{jenkins2014}, another is a version of, “community colleges might not be for you if...” \citep{jenkins2015}, and the latest, \citet{jenkins2018}, discusses why you shouldn't let ``prestige bias'' prevent you from applying to teach at a community college.

If you decide to take the plunge and apply, research that institution/district and make sure the culture seems inviting to you, like a place you could call home. Nearly all the advertised positions will be ridiculously oversubscribed so do what it takes to give yourself an edge and rise above the crowd. Justin Zackal \citeyearpar{zackal2014} has some advice for those with little teaching experience in his article, “Becoming a Community College Professor.” It's a good idea to play up any leadership positions you've previously held, particularly those with a focus on teamwork and community-building. Potential employers want to see that you can motivate folks and nurture a culture of open-mindedness and shared knowledge while still being appropriately authoritative and meeting deadlines. Help them understand that your experience training and leading others has some analogs with facilitating learning in a college classroom. Partnerships and collaborations with other companies and educational institutions and especially outreach experience akin to informal education are also important components. Have you taken any courses in education, even if informal and in a more casual, online environment? Have you read up on developing a personal teaching philosophy? Do you read any newsletters, blogs, or journals on education or teaching? If so, highlight them in ways that help selection committees understand that you are serious about becoming a professional educator.

If you are fortunate enough to get an interview, know what to expect. Given what ought to be seemingly endless variations in the process, it was a little surprising to read in \citet{greenciez2010} just how similar their description of what it should include  is to what actually happens at my own institution. Everyone invited for an interview probably looks like a fantastic teacher and competitive candidate on paper – that’s why they got interviews. \textit{What makes \underline{you} stand out?} Are you connected to teaching communities in your field? Have you done community service or outreach? Hiring a full-time faculty member is a huge investment for that institution. They’re going to want the whole package. If you’re not truly competitive at all those levels, you might lose out. And, most importantly, recognize that even if you ultimately get the position, it’s because the selection committee thought “this person shows great promise,” \underline{not} “this person is an awesome finished product” \citetext{D. Loranz, personal communication, March 21, 2018}.

If you’re interviewing for a part-time position, virtually all of this stuff still applies. Colleges want faculty who are invested in the profession and the best interests of the students, and who are prepared to make a commitment to their institution. Even if there is less competition for the position(s), the ones for whom the profession is part of their identity will rise to the top.

\subsubsection{You're in it now}
\label{initnow}
Congratulations, you got the job! If it's a full-time position remember you may be the only astronomer. You’re probably responsible for all things astronomical at your institution, from course and program development and management to budgets; from recruiting and supervising part-time faculty to starting or maintaining an outreach program; from advocating for facilities and resources to handling inquiries from community members (“I saw this really strange light in the sky last night and it moved really weird. What was it?” or “I found this cool-looking rock in my backyard and it’s got to be a meteorite! Can you verify it for me?”). Juggling all this in combination with your teaching load and institutional responsibilities can be overwhelming at times. It’s important to avoid burnout and especially if you’re new at it, avoid the “teaching trap” \citep{rockquemore2015}. Have a good support structure in place. Cultivating productive relationships with your colleagues in related disciplines is one way to help combat a sense of isolation. It’s also good for your own professional development: visit each other’s classes and schedule regular times to meet up and discuss teaching techniques, the latest research, or new learning technologies. Even the occasional discussion about campus politics, the latest administrative decisions, or just a good old-fashioned bout of commiseration is necessary to keep you grounded and sane. In some cases, it might also mean you have “backup” – others who might be willing and able to help you divide and conquer particularly burdensome duties in a pinch.

Know your craft and take it seriously. A stand-out candidate for a full-time position emphasized this importance in an interview, “I can have even more of an impact here. If you’re good here then they [the students] get even more out of it because the ones coming through the...four-year school with the crazy entrance requirements, they’ll do well anyway. But here, you really have to have your teaching and learning craft down.” \citetext{D. Loranz, personal communication, March 20, 2018}. \textit{We spend nearly 100\% of our time doing what a lot of faculty at four-year institutions try to get out of: teaching.} Immersion into this culture of teaching as a profession isn’t something that just happens. You must be deliberate about it. It isn’t hard, but you do have to make the effort. Start small, by joining a listserv or reading a blog.

If you are a part-timer, look to the experienced ones. Seek them out at your institution(s) and have the important discussions. Consult with the full-timers and if gatherings of all the discipline or department faculty aren’t a regular thing, see if you can change that. There are more opportunities to flex your teaching muscles than you may realize \citep{shropshire2017}. For example, simply visiting each other's classes and having post-observation discussions can lead to some amazing changes, some small and others transformational.

Whether full- or part-time, there are resources for these types of exchanges already in place at most schools. Seek them out and avail yourself of the opportunities. Seek out the astronomers at other schools in your area. Don’t be afraid to reach out; they might be just as keen to connect as you but couldn’t bring themselves to attempt first contact. Resources like \href{https://astronomy101.jpl.nasa.gov/}{CAE}'s Yahoo group \href{https://astronomy101.jpl.nasa.gov/community/}{``Astrolrner''} or your local chapter of \href{http://aapt.org/}{AAPT} may be able to help connect you with astronomy and physics educators in your geographic region if you have trouble searching on your own.

\subsubsection{Student evaluations of teaching}
\label{SET}
I know, I know; I can practically \textit{hear} your eyeballs rolling. As a rule, students’ evaluations of instructors and their teaching methods are inherently flawed and there are mountains of research on this (\citealp{leef2014,starkfreishtat2014,boringetal2016,kelsky2018}; \citealp[and additional references at][]{facultysharesiteSET}). In addition to the “expected” biases, the issue of whether the students are even qualified to judge “good” teaching is a common complaint among instructors. Candid thoughts like the following embody the frustrations of Astro 101 instructors everywhere who are just waiting for the other shoe to drop.

\begin{displayquote}
Aren’t these surveys really just sampling students’ emotions, weighing their feelings about what they think they need against their internalized expectations of what “learning” science is? Faculty end up frustrated and can waste a lot of time and energy explaining to administrators (and even colleagues in other disciplines!) what astronomy really is and constantly justifying what they do in the classroom. Jobs and careers could be on the line here so faculty can feel real pressure to “dumb down” a course and relax expectations, sacrificing pedagogy and genuine learning just to get better student evaluations \citetext{I. Stojimirovic, personal communication, March 23, 2018}.
\end{displayquote}

If this sounds harsh or unreasonable, consider the kinds of questions found on your last round of student surveys and the context – or lack thereof – in which they were answered. Not only is it an almost certainty that every class in every subject surveyed on that campus got \underline{exactly} the same questions, but the students' \textit{interpretation} of those questions is also highly problematic. Students are not trained in pedagogy, classroom management, or evidence-based teaching practices and, frankly, their study ``skills'' are markedly ineffective \citep{roediger2013}, particularly since most of them equate memorization and recall with learning. Most students have never heard of Bloom's taxonomy \citep[and references therein]{heerbloom2012}; \textit{have you?} Let's face it: negative emotions are much more powerful motivators than positive ones, unfortunately. And if we’re being honest, most humans consistently resist making objective decisions, particularly in the face of direct evidence showing their preconceived notions are either incorrect or at least strongly biased by their feelings and emotions \citep{kolbert2017}. Yet the power we give them, in addition to the weight these surveys carry in our tenure reviews and evaluation cycles is, more often than not, rather disproportionate.

That doesn’t mean that these student surveys are totally useless (you could try data mining the written comments for any meaningful morsels). But it does mean you should prepare accordingly. There are things you can do to reduce bias and get meaningful feedback. For example, the Small Group Instructional Diagnosis (SGID) \citep{SGID_UW} is an evidence-based feedback mechanism conducted as focus groups that, at some institutions, can be utilized in place of -- or at least in addition to -- student surveys. My own institution's SGID implementation \citep{SGID_MCC} has been very effective in helping me and my colleagues combat some of the frustration, biases, and futility of student surveys and I cannot recommend the process highly enough. Invest the time and effort. Your job is basically 100\% teaching now so student evaluations and feedback will be a core component of your tenure review or evaluation cycle each time.

\subsubsection{Discipline nuts and bolts}
\label{nuts+bolts}
The word “astronomy” is conspicuously absent from the \textit{\href{http://aapt.org/Resources/tycguidelines.cfm}{AAPT Guidelines for Two-Year College Physics Programs}} \citep{tycguidelines}, it being mentioned only five times in 35 pages. And though a few parts are now a bit dated, it’s still a recommended read for all astronomy, physics, and even physical science instructors, particularly if you are involved in activities like program reviews and advocating for resources, limits on class size, etc. If your department chair and/or dean are not themselves physicists or astronomers it’s definitely worth asking them to read through it as well. (The appendices include a brief history of two-year colleges in the United States and a summary of the various missions of such institutions and so are excellent for anyone looking for quick, entry-level exposures to these topics.)

\subsubsection{Additional perspectives}
\label{addtlperspectives}
Regarding a more general, discipline-independent perspective, there are numerous writings on what it’s like to be a faculty member at a community college. One of the classic references is that of \cite{grubbetal1999} though based on my own experiences and those of several colleagues over the past 10+ years, I would argue that the situation isn’t nearly as glum as presented there. To be fair, the perspective that budgets are growing ever tighter and administrators are still making illogical decisions based on dollars and “butts in seats” hasn’t changed all that much. But there has been a noticeable positive culture shift over the past decade or so and I suspect the proliferation of DBER during this time period, particularly in the STEM fields, is largely responsible.

Indeed, many now sing our praises while cheerleading for graduate students and advisors everywhere to not discount job prospects at community colleges \citep{jenkins2014,jenkins2018,ball2010}. So if you’re considering it, or just wondering whether it may be for you, you may want to spend some time reading through the following articles and those cited in ``\nameref{applying}.''

\cite{Scott2015} says it like it is, diplomatically calling out the elitist attitudes that contribute to Ph.D.s ignoring the community colleges as potential employers. Though not for everyone, he is right to argue that overlooking them may cause you to “miss out on some of the most gratifying and rewarding work in higher education.” Although focused on the humanities, \cite{arteaga2016} gives some striking examples of why “Ph.D.s (and Advisors) Shouldn’t Overlook Community Colleges” while highlighting emerging partner initiatives between two- and four-year schools. Even if there isn’t a formal partnership like any of these in place at your institution, the information in this essay could be helpful for initiating mentorships. In an older piece, \cite{olmstead2001} describes how teaching at a community college became the best thing that ever happened to her. And finally, a broader, more up-to-date encapsulation of how community colleges have evolved over the years and what the future might hold for us can be found in the excerpt by \cite{gill2016} posted on the \href{https://tomprof.stanford.edu/welcome}{Tomorrow’s Professor} website.

\section{Conclusions}
\label{concl}

No matter if you teach part-time or full-time, teaching in a community college is an adventure all its own. In teaching \textit{astronomy} at a community college, your students will be folks of all ages (from high school to post-retirement) with all levels of education and from all backgrounds with an impressive range of experiences under their belts. You'll probably help these students more than you expect to at first, and in ways you might not have anticipated. The challenges and returns are both somewhat surprising and worth it. The experiences in a community college Astro 101 course can be some of the most delightfully bizarre and refreshing of your career and there are few places like it where you can experience so many facets of humanity simultaneously.

Whether you are completely new to teaching or transitioning from teaching at a different type of institution, come with an open mind and an appetite for professional development. Recognize that your community college colleagues expect you to take your craft very seriously and want to help you succeed in it. Expect to spend nearly all of your time working on your courses, grading, and learning about pedagogy, curriculum development, and institutional service through multiple avenues such as serving on committees, visiting and engaging with other professors about each others' teaching, and participating in professional development opportunities. If you are coming to a community college via a more typical ``academic'' pathway (Ph.D.$\rightarrow{}$postdoc$\rightarrow{}$position at a 4-year institution), understand that you will (even if a bit unwittingly at first) trade the ``publish or perish'' motto for an overwhelming sense of commitment to, and desire to do right by, your students. You will join a cohort of professionals who share this mentality, revere this responsibility, and are eager to help each other -- and our students -- navigate this journey. It is, after all, \textit{everyone's} universe...

\section{Acknowledgments}
\label{acknowl}
Special thanks go to a cohort of extraordinary professionals at \href{http://www.miracosta.edu}{MiraCosta College} and countless colleagues everywhere who have helped this community college instructor to better understand her role and develop her craft. Contributing to this work in particular are Philip Blanco, Conrad Hirano, Daniel Loranz, Irena Stojimirovic, the part-time faculty of the \href{https://tiny.cc/astromcc}{MiraCosta College Astronomy Program}, and the generous participants in \href{https://astronomy101.jpl.nasa.gov}{CAE}’s Southwest Regional Teaching Exchange held annually at \href{https://www.miracosta.edu}{MiraCosta College} since 2010.


\clearpage

\section{Biographical Sketch}
\label{biosketch}

Rica Sirbaugh French is a Professor of Astronomy and Physics and directs the \href{http://tiny.cc/astromcc}{Astronomy Program} at \href{http://www.miracosta.edu/}{MiraCosta College}. She spent years researching star clusters and planetary nebulae before realizing she preferred other targets: non-science majors and their instructors. With collaborators at the \href{https://astronomy101.jpl.nasa.gov/}{Center for Astronomy Education} (CAE) she is part of the nation’s largest college-level astronomy education research initiative and facilitates professional development experiences for educators nationwide. She has served on the board of the North County Higher Education Alliance, managed a professional development program for the California Community Colleges Chancellor’s Office, and maintains resources for faculty at \url{https://tiny.cc/rfrenchfacultyshare}. A member of the \href{http://aapt.org/}{American Association of Physics Teachers} (AAPT), the \href{https://www.astrosociety.org/}{Astronomical Society of the Pacific} (ASP), and the \href{https://www.iau.org/}{International Astronomical Union} (IAU), she is also an Agent and Career Advisor with the \href{https://aas.org/}{American Astronomical Society} (AAS) having served on their Astronomy Education Board, as a panelist for the Committee on Employment, and a columnist for the education newsletter \textit{Spark}.

Originally from Tennessee, Rica and her husband still take their motorcycles to the track, love pro football, and watch too much television. She also enjoys most sports, pretends she is athletic, plays several musical instruments mediocrely, and has recently rediscovered reading for \textit{pleasure(!)}. They live in San Diego County, California.

\end{document}